\documentclass{aa}
\usepackage{graphicx}
\usepackage[varg]{txfonts}
\begin{document}

\title{Detailed photometric analysis of young star groups in the galaxy NGC~300}

\author{M. J. Rodr\'\i{}guez \inst{1}
\and G. Baume \inst{1,2}
\and C. Feinstein \inst{1,2}}

\institute{ Instituto de Astrof\'\i{}sica de La Plata (CONICET-UNLP), Paseo del bosque S/N, 
La Plata (B1900FWA), Argentina
\and Facultad de Ciencias Astron\'omicas y Geof\'\i{}sicas - Universidad Nacional de La Plata, Paseo del bosque S/N, 
La Plata (B1900FWA), Argentina}

\abstract {} {The purpose of this work is to understand the global characteristics of the stellar populations in NGC~300.
In particular, we focused our attention on searching young star groups and study their hierarchical organization.
The proximity and orientation of this Sculptor Group galaxy
make it an ideal candidate for this study.} 
{The research was conducted using archival point spread function
(PSF) fitting photometry measured from images in multiple
bands obtained with the Advanced Camera for Surveys of the Hubble Space Telescope (ACS/HST).
Using the path linkage criterion (PLC),
we cataloged young star groups  and analyzed them from the observation of individual stars in the galaxy NGC~300.
We also built stellar density maps from the bluest stars and applied the SExtractor code to identify overdensities. This method provided an additional tool for the detection of young stellar structures. 
By plotting isocontours over the density maps and comparing the
two methods, we could 
infer and delineate the hierarchical structure of the blue population in the galaxy.
 
For each region of a detected young star group, we estimated the size and derived the radial surface density profiles for stellar
populations of different color (blue and red).  
A statistical decontamination of field stars was performed  for each region. In this way it was possible to build
the color-magnitude diagrams (CMD) and compare them with theoretical evolutionary models.
We also constrained the present-day mass function (PDMF) per group by estimating a value for its slope. } 
{The blue population  distribution in NGC~300 clearly follows the spiral arms of the galaxy,
 showing a hierarchical behavior in which the larger and loosely distributed structures split into more compact and denser ones
 over several density levels.
We created a catalog of 1147 young star groups in six fields of the galaxy NGC~300,
in which we present their fundamental characteristics.
The mean and the mode radius values obtained from the size distribution are both 25 pc,  
in agreement with the value for the Local Group and nearby galaxies.
Additionally, we found an average PDMF slope that is compatible with the Salpeter value.
} {}


\maketitle

\section{Introduction}

The distribution and properties of stellar clusters and star-forming complexes provide valuable
information
that helps understanding the history of the star formation inside galaxies
and the way this is related with the dynamic and chemical enrichment of each system. 
This reveals the importance of performing global studies and assembling catalogs of clusters
over the nearest galaxies.  

The galaxy NGC~300 is the brightest of five main spiral galaxies forming the Sculptor Group.
It is a spiral galaxy Sc D \citep{2012AJ....144....4M} and presents several regions of 
massive star formation. 
This galaxy also has an orientation that minimizes the absorption effects, and it is located 
close enough to identify stellar clusters and their individual
members by using images from telescopes with excellent angular resolution.
In particular, \citet{2001A&A...371..497P}  have provided a catalog with over
one hundred OB associations in this galaxy using the path linkage
criterion (PLC) and ground-based observations
obtained with the ESO/MPI 2.2 m telescope. 

The aim of this work is to catalog and analyze the young star groups 
in NGC~300 using excellent quality data from the Hubble Space Telescope (HST).
In particular, the Advanced Camera for Surveys (ACS) provides the quality 
we require, and some of the observed fields cover an important 
part of the galaxy NGC~300 \citep{2005ApJ...634.1020B}. 
These data will allow us to obtain much more detailed information for the 
stellar groups than ground-based observations, such as the corresponding radial profiles
and photometric diagrams.
All these tools and their comparison with theoretical models 
will help us understand the properties of their individual populations.

The paper is organized as follows:
In Sect. \ref{data} we describe the observations and data sets.
The reduction, astrometric corrections, and correlation procedure
of the data are presented in Sect. \ref{reduction}.
Section \ref{search} discusses the method we used to search for and identify stellar groups.
Section \ref{analysis} presents our analysis of the detected associations. 
We discuss our results concerning the size, mass function, and hierarchical behavior in Sect. \ref{discussion}.
Finally, in Sect.  \ref{conclusions} we summarize our results.

\section{Data and observation}
\label{data}

\begin{figure}
\includegraphics[scale=0.15]{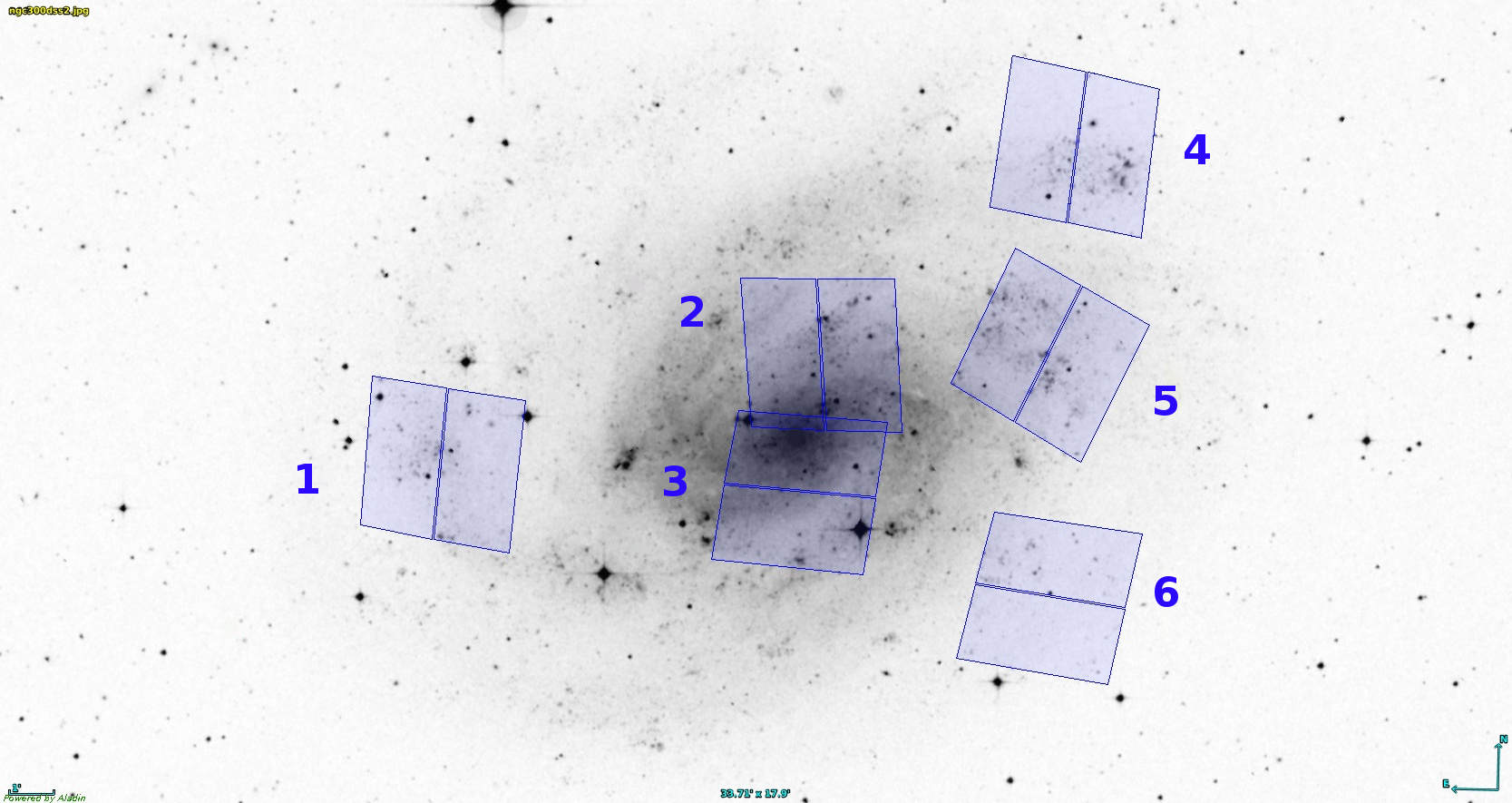}
\caption{Distribution of the different fields used in this work 
 overlaid on a DSS image of NGC~300.}
\label{campos}
\end{figure}

The data used in this work, images, and photometric tables correspond
to the "star files" from the ACS Nearby Galaxy Survey (ANGST).
They were obtained from the database of the Space Telescope Science Institute (STScI)
\footnote{MAST: https://archive.stsci.edu/; https://hla.stsci.edu/}.
These files contain the photometry of all
objects classified as stars with good signal-to-noise values (S/N > 4) 
and data flag <~8.
The observations were carried out with the camera ACS/HST.
They correspond to six fields in NGC~300 (see Fig.\ref{campos})
and were obtained
during HST Cycle 11 as part of program GO-9492 (PI: F. Bresolin).
These observations employed the Wide Field Camera (WFC) 
of the ACS, obtaining exposures of 360 sec., three in the F435W and
F555W bands and four exposures in the F814W band, over each field.
The WFC has a mosaic of two CCD detectors and a scale of 0.049 $"$/pixel,
the field of view covered is 3.3' x 3.3'.

At the considered distance of 1.93 Mpc \citep{2005ApJ...634.1020B},
the following relationship applies: 1" = 20 pix $\sim$ 9.4 pc.

\section{Photometry and astrometry}
\label{reduction}
\subsection{Photometry}

The binary fits tables of photometry as defined in \citet{2008glv..book..115D}
were obtained from the STScI data base.
The high image resolution allows stellar point spread function (PSF) photometry 
in nearby galaxies. 
This enables studying their different stellar populations in great detail.
The photometry was carried out using the package DOLPHOT adapted 
for the ACS camera \citep{2000PASP..112.1383D}.

\subsection{Astrometry}

Using the ALADIN \footnote{http://aladin.u-strasbg.fr/} tool to compare the images 
of each observed NGC~300 field with the respective \citet{2008glv..book..115D} catalog,
a systematic difference (see Table~\ref{table:1}) was found between the World Coordinate System
(WCS) indicated in the image headers and those given by the catalog.
Then, we compared the positions with the astrometric catalogs UCAC
4 and GSC2.3 and verified that the WCS coordinates were correct.
Next we accurately evaluated the difference between the two coordinate systems.
To do this, we used the task xy2rd of the IRAF's \footnote{IRAF is distributed by NOAO,
which is operated by AURA under cooperative agreement with the NSF.}
stsdas package to obtain 
the WCS coordinates of some (see Table ~\ref{table:1}) bright and unsaturated stars
in the F555W band images of each field. 
We compared these values with the corresponding ones 
given by \citet{2008glv..book..115D}. 
These corrections were applied to all the objects in this catalog.
Finally, we compared each resulting catalog with its 
corresponding image using ALADIN. 
The catalogs and images agreed well,
although we note that one small relative distortion ($\sim 0.1"$)
still prevailed over some fields.

\begin{table}
\caption{Corrections applied to the coordinates given by \citet{2008glv..book..115D}.
In Col. 1 we indicate the frame ID, the corresponding corrections are given in 
Cols. 2 and 3. Column 4 lists the number of stars used in the coordinate alignment procedure. 
}             
\label{table:1}      
\centering          
\begin{tabular}{c c c c}     % 4 columns 
\hline\hline       
                      
Frame & $\Delta\alpha \cos(\delta) ["]$ & $\Delta\delta ["] $ & N\\ 
\hline 

  1   &    $3.283\pm0.024 $ &  $-1.434\pm 0.044$ & 5 \\
  2   &    $0.335\pm0.021 $ &  $-2.316\pm 0.087$ & 3 \\
  3   &    $-1.247\pm0.032$ &  $-0.462\pm 0.050$ & 3 \\
  4   &    $ 0.519\pm0.027$ &  $-2.541\pm 0.078$ & 3 \\
  5   &    $-0.355\pm0.018$ &  $0.886\pm 0.007$  & 6 \\
  6   &    $-0.021\pm0.004$ &  $0.029\pm 0.009$  & 7 \\
\hline                  
\end{tabular}
\end{table}

\subsection{Catalog correlation}
\label{catalog}

The catalogs given by \citet{2008glv..book..115D} provide photometric information
through three tables for each field, giving in each one
the magnitudes of only two bands for each object.
We therefore constructed a catalog with the three photometry bands for each field from them.
This was performed with the STILTS \footnote{http://www.star.bris.ac.uk/~mbt/stilts/}
cross-correlation (with logical "OR") between
tables with $F555W-F435W$ bands and those with $F814W-F555W$ bands.

Furthermore, fields 2 and 3 slightly overlap (see Fig. \ref{campos}),
therefore  we took a final magnitude for the stars in the overlap region that results from the average
between the stars in each field.  Using STILTS, we then joined the information and list the two fields as one single field in
the table.

\section{Searching for young stellar groups}
\label{search}
To evaluate the completeness of the data, we built the luminosity function for each table (see Fig. 2).
The number of stars per bin begins to decrease below a magnitude $F555W=$ 27. Therefore we considered that 
the sample is complete for $F555W <$ 26.5.

\begin{figure}
\resizebox{\hsize}{!}{\includegraphics{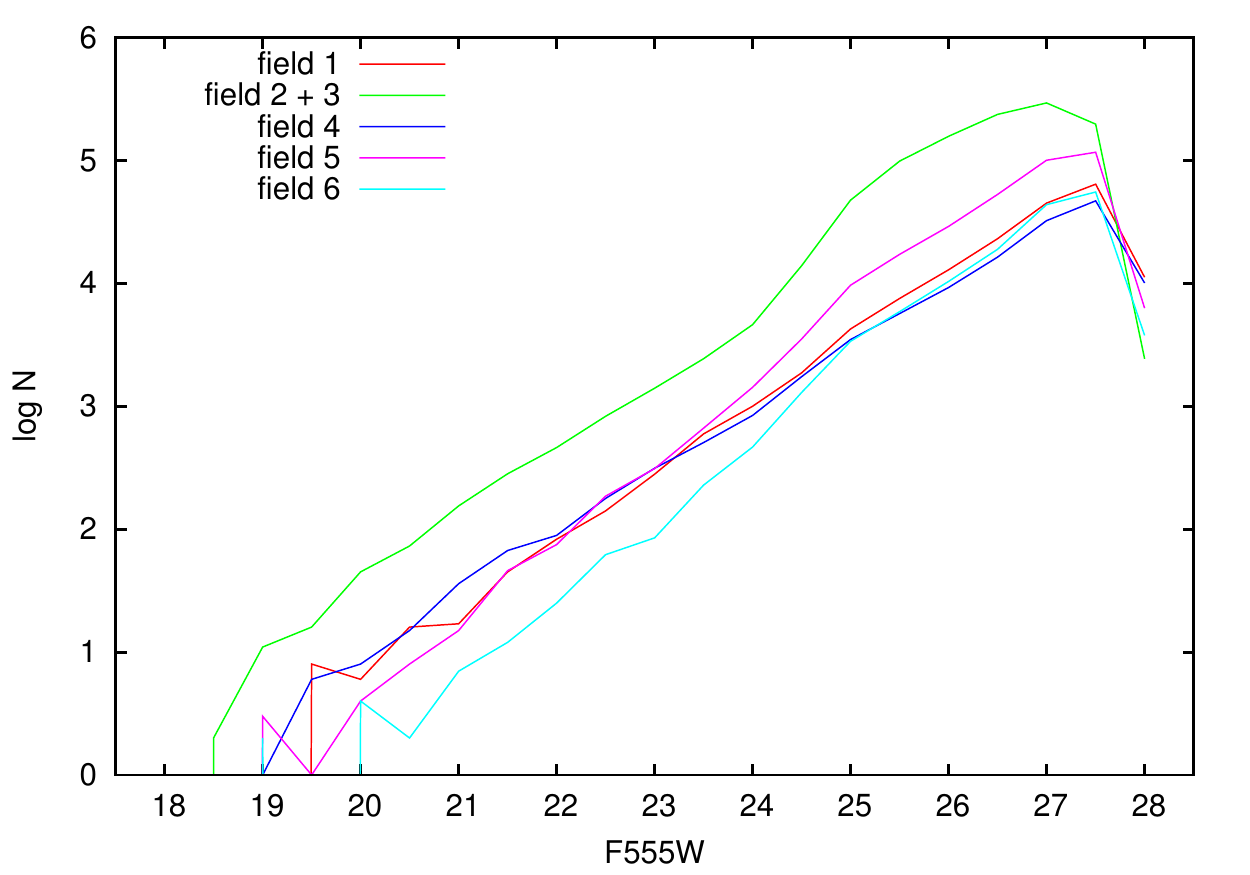}}
\caption{Luminosity functions for the studied fields (fields 2 and 3 are reported as one single field).} 
\label{FL}
\end{figure}

For better identification throughout the text, we distinguish the following four groups of stars:
\begin{itemize}
 \item the blue group: $F435W-F555W < 0.25$ and $F555W-F814W < 0.25$;
\item the blue bright group: $F555W < 25$, $F435W-F555W < 0.25$ and $F555W-F814W < 0.25$;
\item the red group: $F435W-F555W > 0.6$ and $F555W-F814W > 0.6$;
\item the red bright group: $F555W < 25$, $F435W-F555W > 0.6$ and $F555W-F814W > 0.6$.
\end{itemize}
In Fig. 3 we show the CMDs of all the stars in the six fields, indicating the previous groups 
of stars and the location of particular isochrones for solar metallicity \citep{2008A&A...482..883M,2010ApJ...724.1030G}.
For the blue bright sample at F555W = 25, the oldest
main-sequence turn-off corresponds to 235 Myr, and the lowest stellar mass value for a  main-sequence 
star is $\sim 7.4 M_{\odot}$.

\begin{figure}
\resizebox{\hsize}{!}{\includegraphics{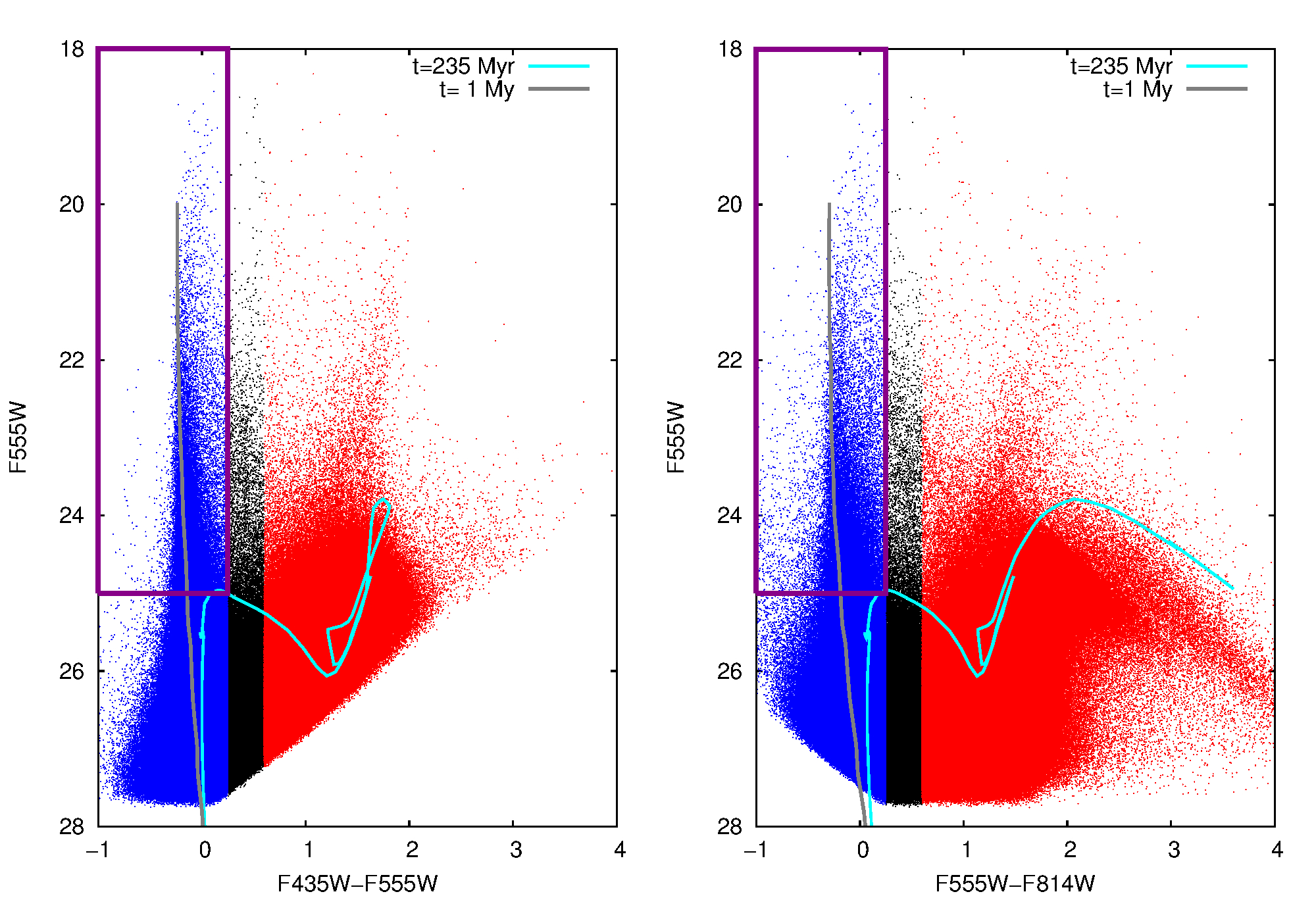}}
\caption{Total CMDs of the six NGC 300 fields with indicative isochrones 
with solar metallicity (see Sect. \ref{masses})
corresponding to 1 Myr (grey line) and 235 Myr (turquoise line). Inside the 
rectangle we plot the stars belonging to the blue bright group;
red and blue show the red  and blue group. }
\label{cmd-t}
\end{figure}

We used the selected blue bright group to detect young stellar structures, in particular
OB associations and young stellar clusters. For this task, we applied different methods that
we describe in Sects. \ref{density} and \ref{secPLC}

\subsection{Stellar density maps}
\label{density}

With the aim of better understanding the distribution of different stellar populations in the galaxy
and to have an additional tool to identify young star groups,
we used the blue bright group and the red bright group to build spatial density stellar maps for the observed regions.
In this process, we constructed for each field a two-dimensional histogram counting
the numbers of stars in spatial bin sizes of 8.0 arcsec. Then we applied the drizzle 
method \citep{2002PASP..114..144F} considering a 2.0 arcsec step.
The obtained density maps (six) were combined into one image with astrometrical calibration. 

Then, we used the Source Extractor code (SExtractor)\footnote{http://sextractor.sourceforge.net/} 
on the obtained blue stellar density map to identify overdensities that could be identified as young
stellar groups. 
To do this we adopted the default convolution mask of two pixels and
assumed a size of four pixels for the background mesh.
A larger mesh would imply that the background of the density stellar maps
would be too smooth, and therefore some sources might be lost.
A detection threshold of 1.5 $\sigma$ above the local background was adopted, 
and objects with more than four pixels were considered as reliable overdensities.
We assumed a value for the deblending contrast parameter $\delta_{c}$=0.001
because this gives the best separation between sources.

The application of this method resulted in the detection of 
289 candidates for young groups.

\subsection{Path linkage criterion}
\label{secPLC}

We also employed the PLC technique \citep{1991A&A...244...69B} 
to identify young star groups in NGC 300. 
This method consists of connecting blue stars that are less distant 
than a fixed parameter called $d_{s}$.
When it is possible to link more than $p$ stars, we have a candidate OB association or stellar clusters. 

\begin{figure}
\resizebox{\hsize}{!}{\includegraphics{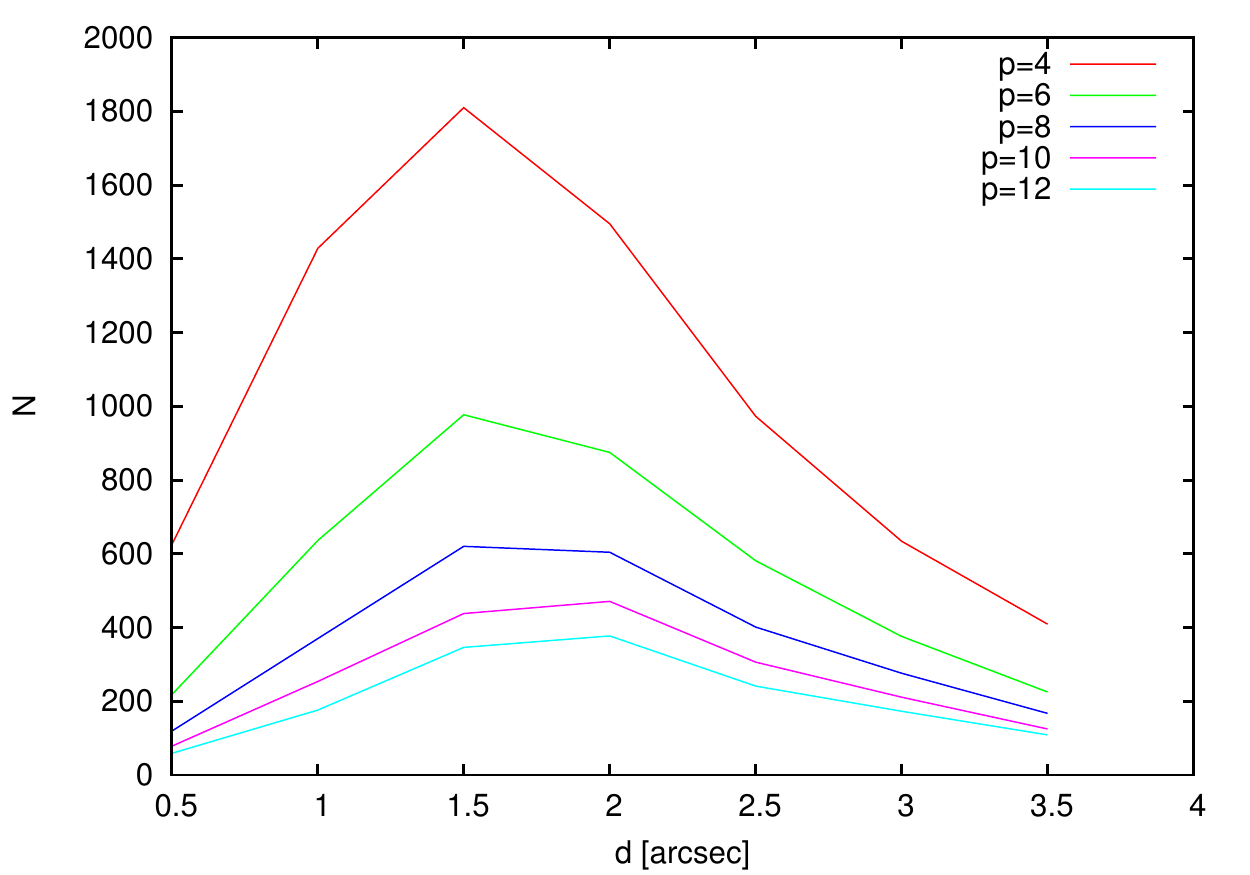}}
\caption{Behavior of the number of groups detected using the PLC technique with respect to the parameter $d_{s}$
for different values of the parameter $p.$}
\label{plc}
\end{figure}

We applied this method to the selected blue bright group 
of stars on each covered field.
Then, we had to choose adequate values for the parameters $p$ and $d_{s}$.
The lowest value of the number of stars that a group should have to be considered
an association or stellar cluster ($p$) is a parameter difficult to establish, since a low value results
in many spurious detections, while with a high value the smallest groups might be lost.
We studied the number of groups identified using PLC with respect to the parameter $d_{s}$
for different values of $p$ (see Fig. \ref{plc}). 
Finally, we considered that ten stars is a reasonable value for the parameter $p$ and adopted a value range between 0.5-2.5 arcsec for $d_{s}$.
The advantage of starting the search with a small distance is to detect the 
smallest subgroups of each large association or stellar complex that otherwise
would have been lost in larger structures.
After detecting the smallest groups, 
the star members of these groups were eliminated from the sample and the PLC 
was run again, but increasing the value of $d_{s}$. This procedure was repeated until 
$d_{s}$ reached 2.5 arcsec. Using this method, we detected the individual groups but not the form 
in which they merge into larger structures. We discuss this point in Sect. \ref{hierarchical}.  

Using the PLC technique, we detected 1147 young stellar groups in the six fields. 

\subsection{Results of the search}

\begin{figure*}
\centering
\includegraphics[scale=0.4]{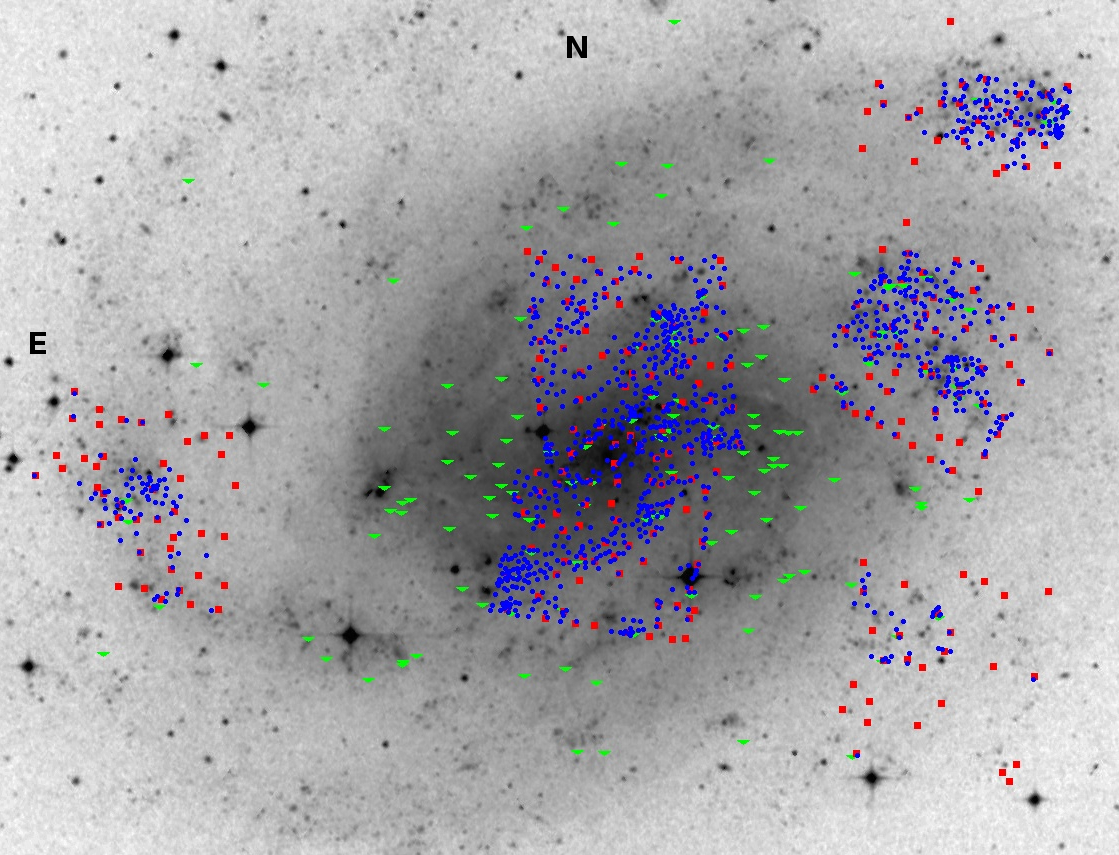}
\caption{Results obtained using SExtractor (red squares), PLC (blue circles), 
and the catalog obtained by \citet{2001A&A...371..497P} (green triangles)
overlaid on a DSS image of NGC 300.}
\label{catalogos}
\end{figure*}

\begin{figure*}
\centering
\includegraphics[width=17cm]{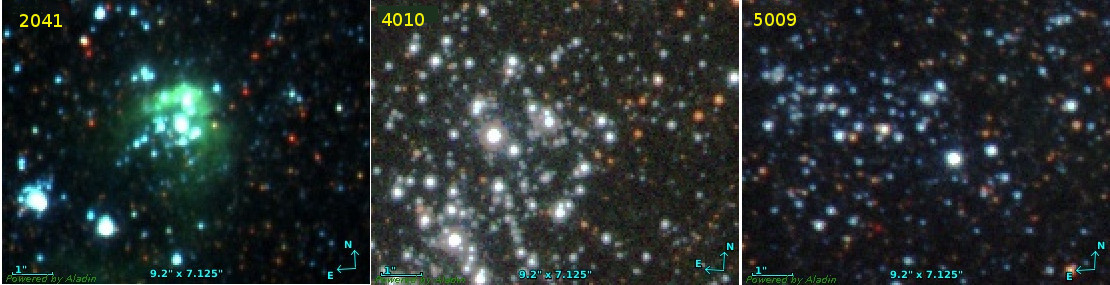}
\caption{Color images of the associations 2041, 4010, and 5009 located at different 
galactocentric distances. These groups were  detected with the PLC method.}
\label{associations}
\end{figure*}

All the OB associations detected by \citet{2001A&A...371..497P} and located in the six ACS fields
were found with both techniques used in this work.
The obtained detections with the different techniques and the results obtained by \citet{2001A&A...371..497P} 
are shown in Fig. \ref{catalogos}, in which we can see from the PLC results that the young stellar groups
clearly delineate the spiral structure of the galaxy. 
The associations identified using the SExtractor code show a more sparse distribution because the group sizes are larger, since the 
pixels of the density maps are four times larger than the ACS/WFC. This causes several PLC 
detections to correspond to a single detection of SExtractor. This is a sign of the hierarchical distribution of the blue population in 
the galaxy, which we address in our discussion (see Sect.~\ref{hierarchical}). For this reason and considering that we identified
some spurious detections as border effects with SExtractor, we decided to adopt the PLC detections as the final catalog of young stars groups in the galaxy.
In Table \ref{tabla2} we show the first ten rows of the catalog, the complete version is available online.

\begin{figure}
{\includegraphics[height=12cm]{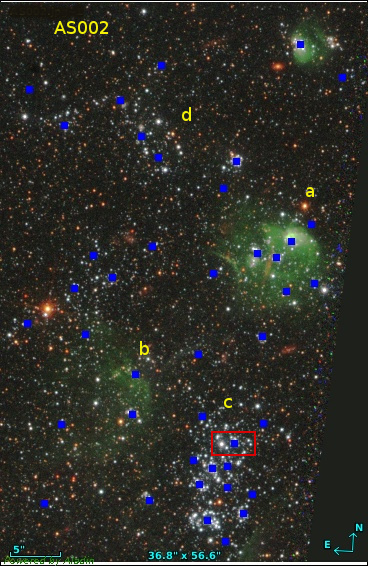}}
\caption{Stellar complex AS002 named by \cite{2001A&A...371..497P}. The letters a to d designate the four subgroups detected in their work. Our PLC
detections are indicated with blue squares. The red rectangle encloses the association 4010, which is shown in 
Fig. \ref{associations}.}
\label{AS2}
\end{figure}

In Fig. \ref{associations} we show three PLC-detected associations at different galactocentric distances. In particular, association 4010 is located 
in the stellar complex AS002 named by \cite{2001A&A...371..497P}. As we show in Fig. \ref{AS2}, they detected four subgroups (a, b, c, and d). 
Using the ASC/HST observations and the PLC method, we detected at least 40 groups in the same region, which are indicated with blue squares in the figure.

\section{Analysis}
\label{analysis}

To perform a systematic, homogeneous, and efficient analysis of each
identified group, we developed a dedicated numerical code in FORTRAN 95.

\subsection{Size and radial density profile}

The coordinates of the center and the sizes of the groups were estimated as 
the mean values and the dispersions of
the positions of the blue bright group stars identified for each association.

\begin{figure}
\resizebox{\hsize}{!}{\includegraphics{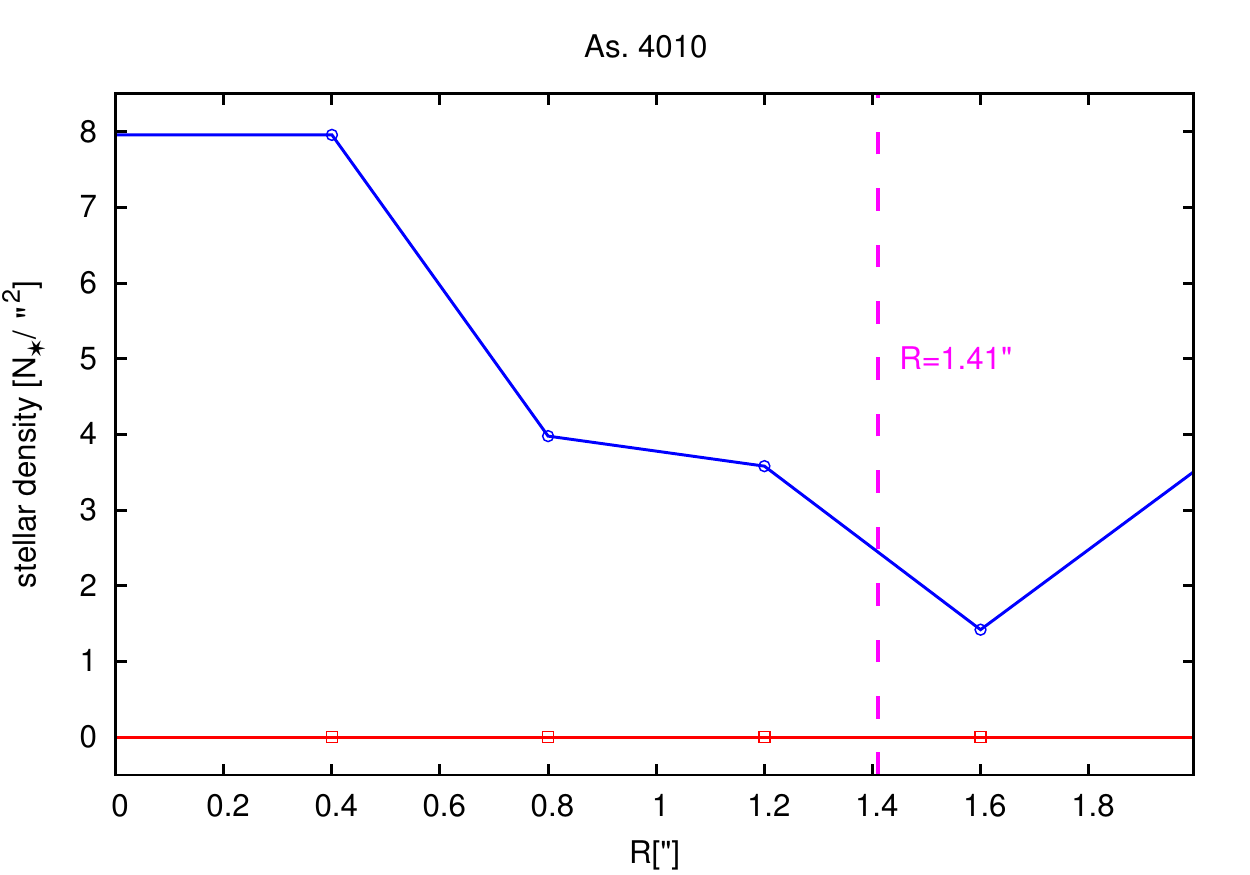}}
\caption{Radial density profile of association 4010. With blue and red we indicate 
the density of blue  and red group stars; 
the pink dashed line shows the adopted association radius at 1.41".}
\label{PR-4010}
\end{figure}

The density profile was built by distinguishing the different color populations;
these are the blue and red group. 
Then, the stellar density for each color was measured, considering
concentric rings with a radial step of 0.4 arcsec.
In Fig. \ref{PR-4010} we show the radial density profile of association
4010. A clear overdensity of blue stars is evident, while the density of red stars
remains equal to zero. This is expected for a young stellar association.

\subsection{Field star decontamination}

Our code uses the coordinates of the previously localized groups and performs a statistical 
decontamination of the field stars based on the magnitudes and color indexes of the stars 
\citep[see][]{2003AJ....125..742G}. 
The procedure consists of a star-by-star comparison of the quantities $F555W$,
$F435W-F555W$ and $F555W-F814W$ between the stars of the group and the stars of a field region
located near the association and covering the same sky area.
Then, we subtracted the stars in the group that matched  a star in the field region in these
quantities.
To obtain results as accurate as possible, the code carries out this procedure with five different 
field regions for each association. The first region is a ring around the association, 
and the others are selected as follows: ($\alpha_{0}\pm\Delta\alpha $, $\delta_{0}$);
($\alpha_{0}$, $\delta_{0}\pm\Delta\delta$) where $\Delta\alpha=\Delta\delta=20"$.
We discarded the field regions with the largest and smallest number of stars to avoid
possible neighboring associations or to fall outside the field of view. 
The final parameters of an association are an average of the parameters obtained with the three remaining field regions.

From these results we built the photometric diagrams and mass histograms
and calculated the fundamental parameters of each group in this
way.

\subsection{Color-magnitude diagrams}

\begin{figure}
{\includegraphics[scale=0.65]{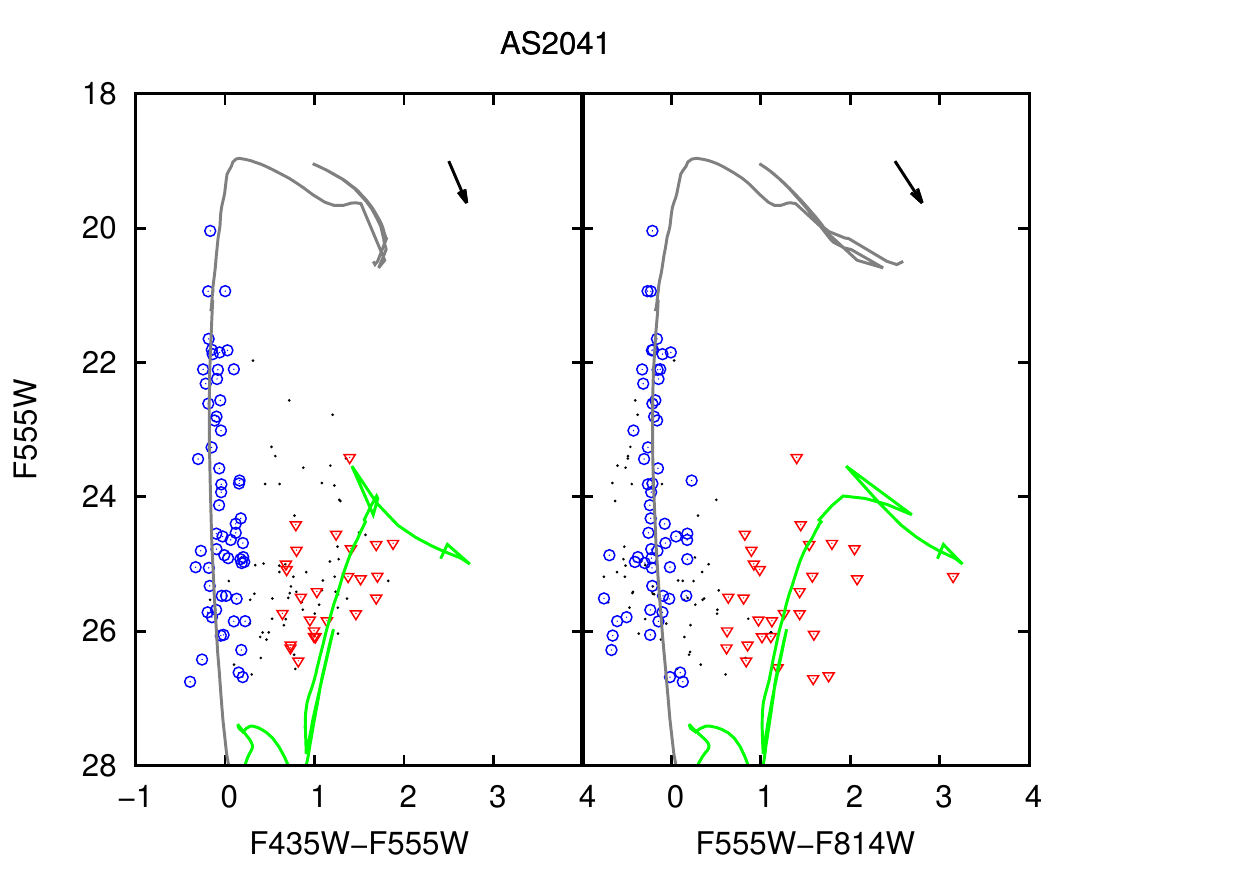}} \\
{\includegraphics[scale=0.65]{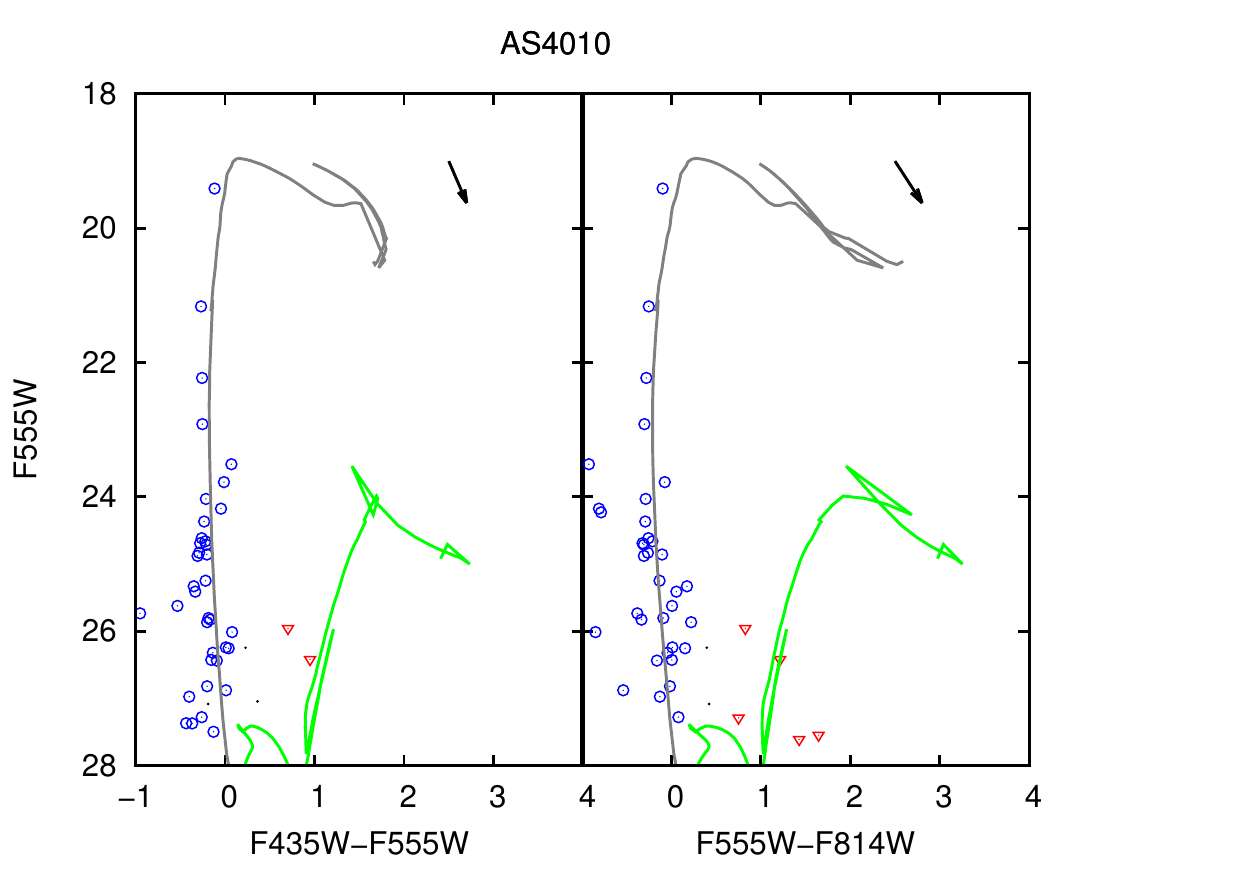}} \\
{\includegraphics[scale=0.65]{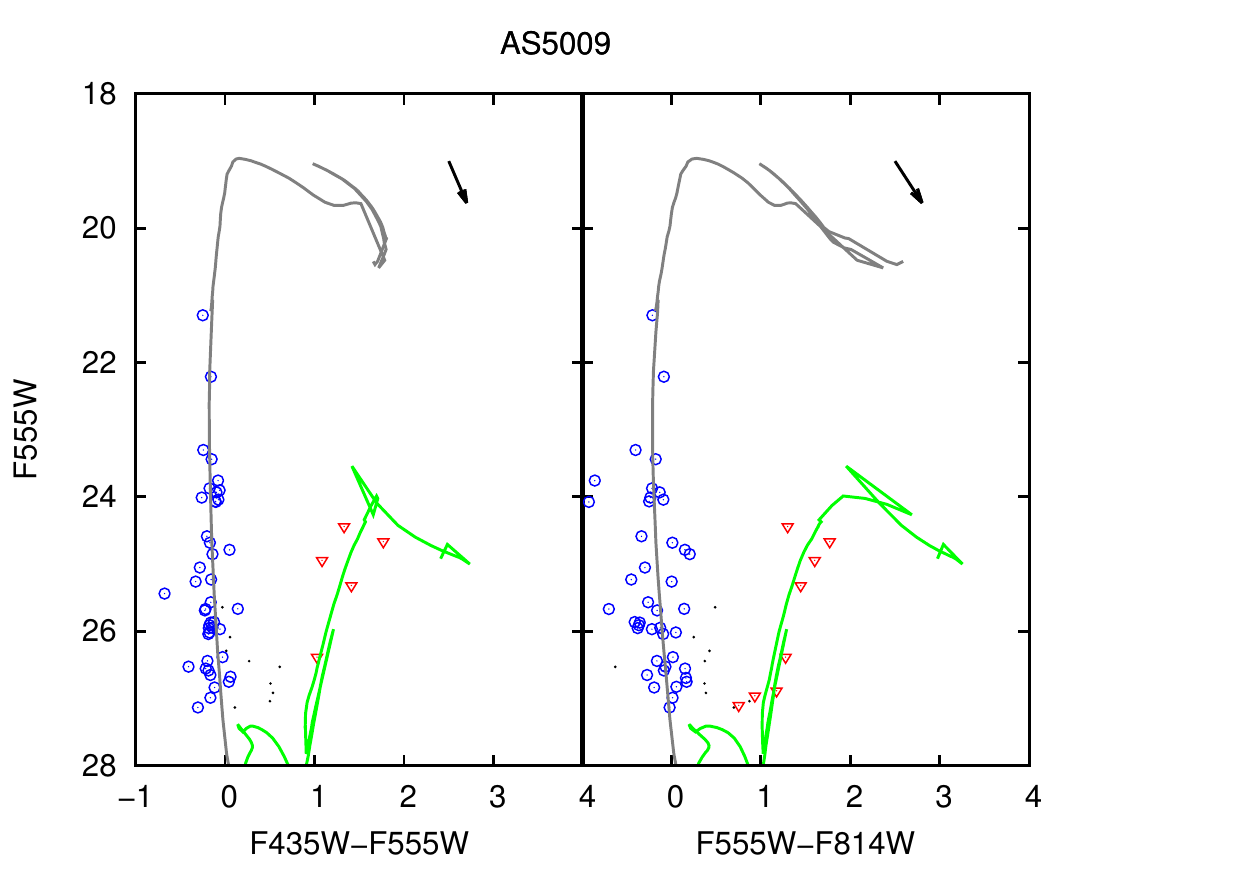}} \\
\caption{Decontaminated color-magnitude diagrams $F555W$ vs. $F435W-F555W$
and $F555W$ vs. $F555W-F814W$ of the stellar associations
2041, 4010, and 5009
that are located in different places of the galaxy.
The blue circles represent
the blue group of stars, the red inverted triangle indicate the red group, 
and the black dots are the stars that are neither blue nor red.
The gray line indicates the isochrone corresponding to $10^7$ yr 
and solar metallicity ($Z$=0.019). The green line corresponds to $10^9$ yr
and $Z$=0.008 \citep{2008A&A...482..883M,2010ApJ...724.1030G}.
The arrow indicates the reddening vector.}
\label{CMDs}
\end{figure}

We used the blue and red group to build the color-magnitude diagrams (CMDs) 
shown in Fig. \ref{CMDs}. The  evolutionary models of \citet{2008A&A...482..883M}
with the \citet{2010ApJ...724.1030G} corrections, corresponding to different 
ages and metallicities, were superposed for comparison.
To displace the models and compare them with our data,
we adopted a distance modulus of $(Vo-M_{V})=26.43$ \citep{2005ApJ...634.1020B},
a normal reddening law ($R =A_{V}/E(B-V)$= 3.1),
and a value for $E(B-V)$=0.075, where the Galactic foreground 
reddening toward NGC 300 of 0.025 mag and an additional reddening of
$E(B-V)$=0.05 mag intrinsic to NGC 300 were considered \citep{2004AJ....128.1167G}.

The obtained CMDs for most of the identified young star groups revealed a sharp
distribution of blue stars. This fact was understood as a consequence of a
very low reddening inside the galaxy \citep{2004AJ....128.1167G}. 
This situation eliminates possible degeneracy between temperature and extinction for hot
stars. Then, as in the following section, we were able to estimate stellar masses from 
their magnitudes and compare them with fitted theoretical evolutionary models.

\subsection{Masses and the present-day mass function}
\label{masses}

\begin{figure}
{\includegraphics[scale=0.6]{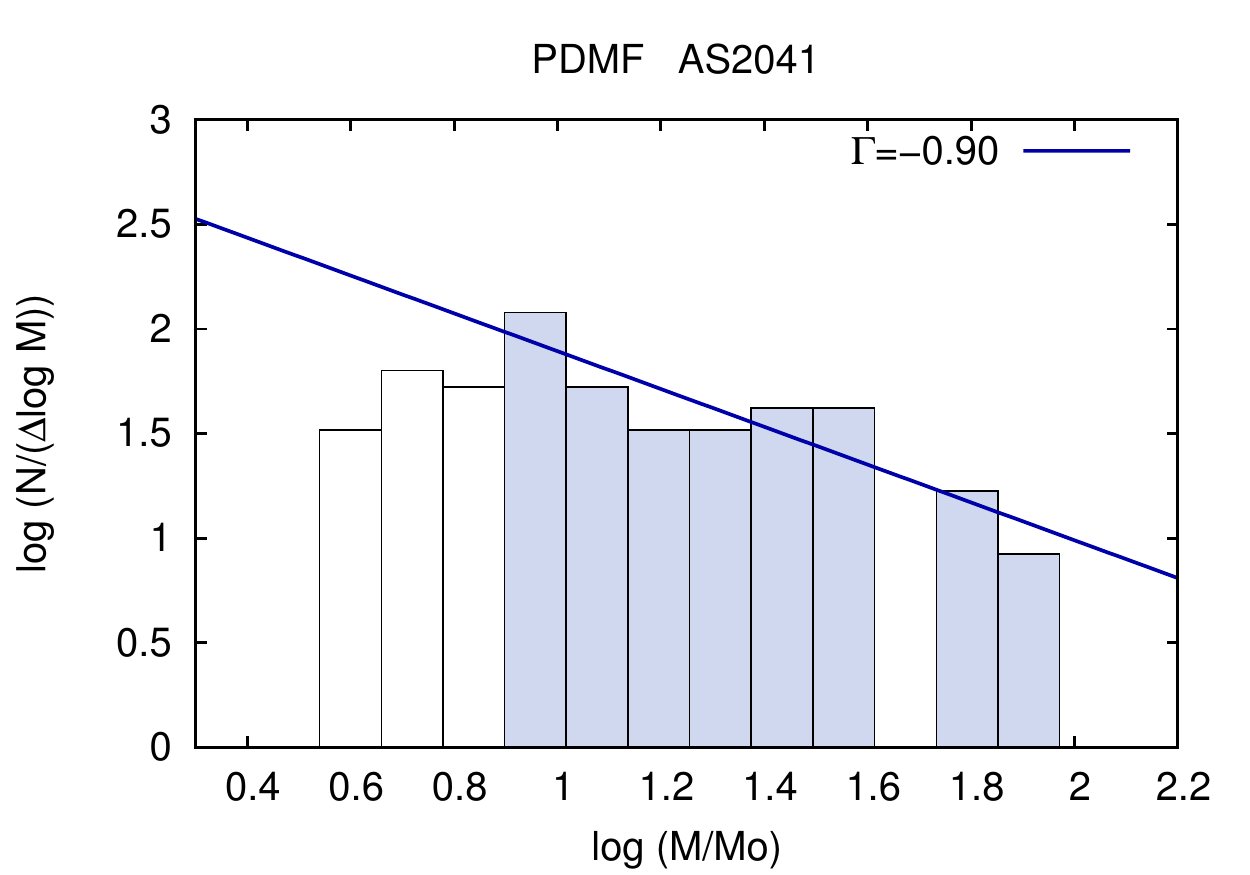}} \\
{\includegraphics[scale=0.6]{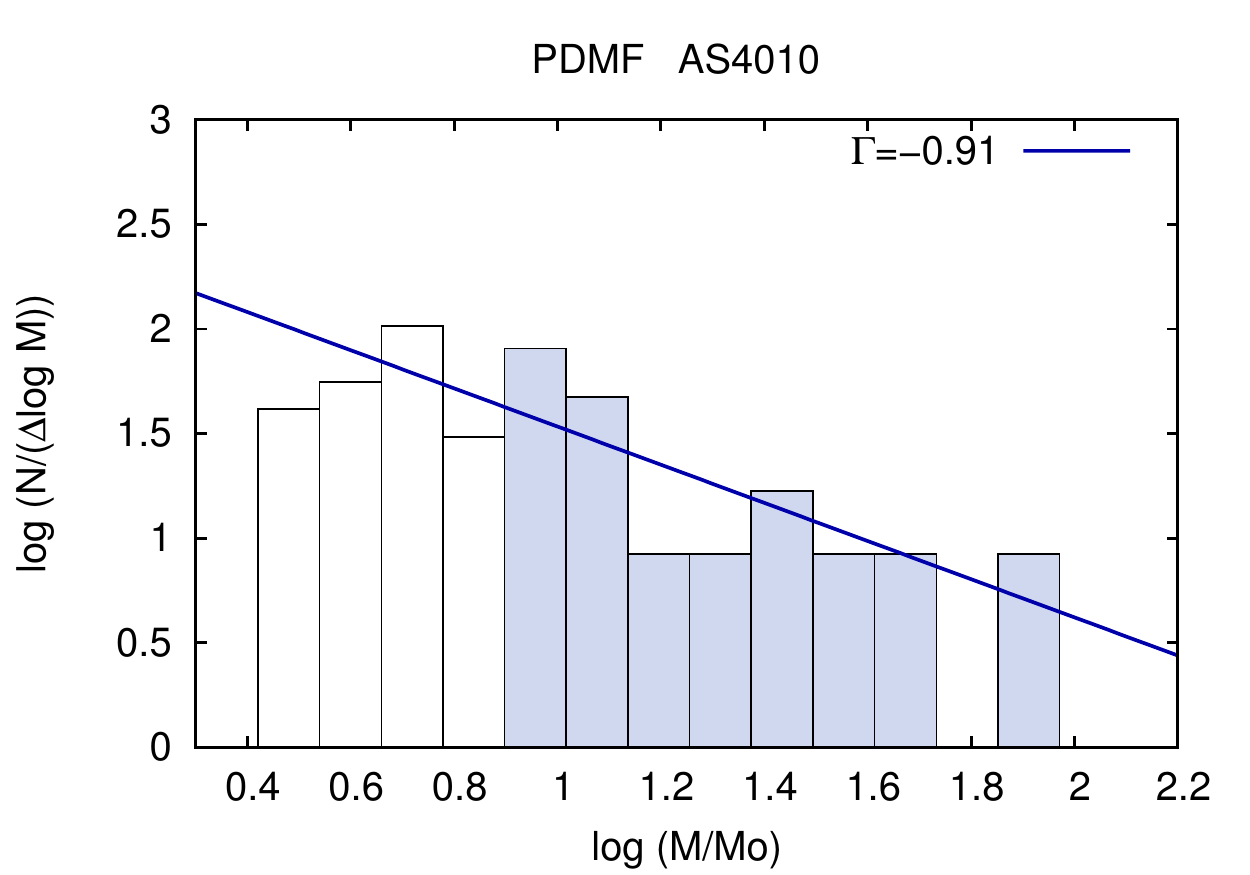}} \\
{\includegraphics[scale=0.6]{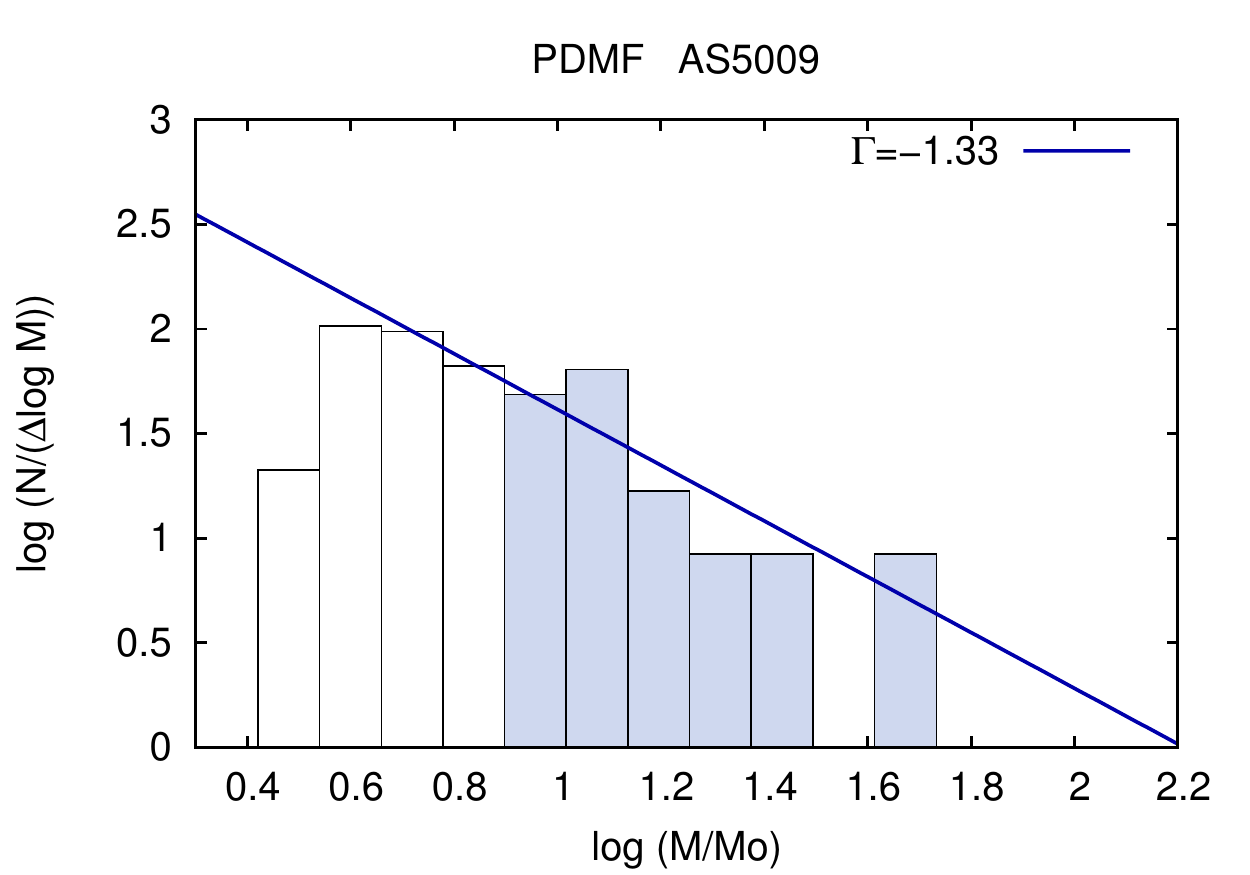}} \\
\caption{Present-day mass function of the stellar associations
2041, 4010, and 5009 that are located in different places of the galaxy.
The line corresponds to a linear fit over the gray bins (M $>$ 7.4 M$\odot$).}
\label{IMF}
\end{figure}

With the purpose of deriving stellar masses, 
we performed
a lineal interpolation in the F555W band 
in the adopted evolutionary models of 1 Myr (see Sect. \ref{search}).
To avoid foreground stars in each association, we estimated the
stellar masses of the blue-group stars resulting from the statistical decontamination.
Given that NGC 300 presents an abundance gradient with Z between 0.004-0.018
\citep {2009ApJ...700..309B, 2015ApJ...805..182G},
we tried to estimate the masses with a range of metallicities between $Z=0.007$ and $Z=0.019$.
Finally, we adopted solar metallicity, since we did not obtain significant differences. 
The corresponding obtained mass histograms are presented in Fig. \ref{IMF}. 
Then, using only those bins corresponding to the high-mass range 
(M $> 7.4~M_{\odot}$, which corresponds to the blue bright group),
we performed linear fits to the data. 
These histograms are representative of the corresponding present-day
mass functions (PDMFs). Additionally, in the adopted mass range, the PDMF can be 
modeled as a power law and expressed in the form
$log(N/\Delta (log m)) = \Gamma \Delta(log m),$
where N is the number of stars per logarithmic mass bin log(m).
The computed corresponding slope values ($\Gamma$) are also indicated in Fig. \ref{IMF},
and the obtained PDMF slopes for each detected association are presented in Table \ref{tabla2}. 

To estimate  the effect of galaxy distance uncertainty and/or possible variable extinction in
our procedure, we repeated it for different distance values ($26.43\pm 0.09$, \cite{2005ApJ...634.1020B})
and using $A_V$ values ranging from 0 to 0.53 \citep{2005ApJ...632..227R}.
We obtained that the changes in stellar mass are $\Delta M / M \sim 25\%$
and changes in the corresponding PDMF slopes are lower than the indicated fitted errors.
Therefore, our analysis and conclusions are independent of the above indicated effects.

\section{Discussion: Global properties of the young star groups}
\label{discussion}
\subsection{Size distribution}

\begin{figure}
\resizebox{\hsize}{!}{\includegraphics{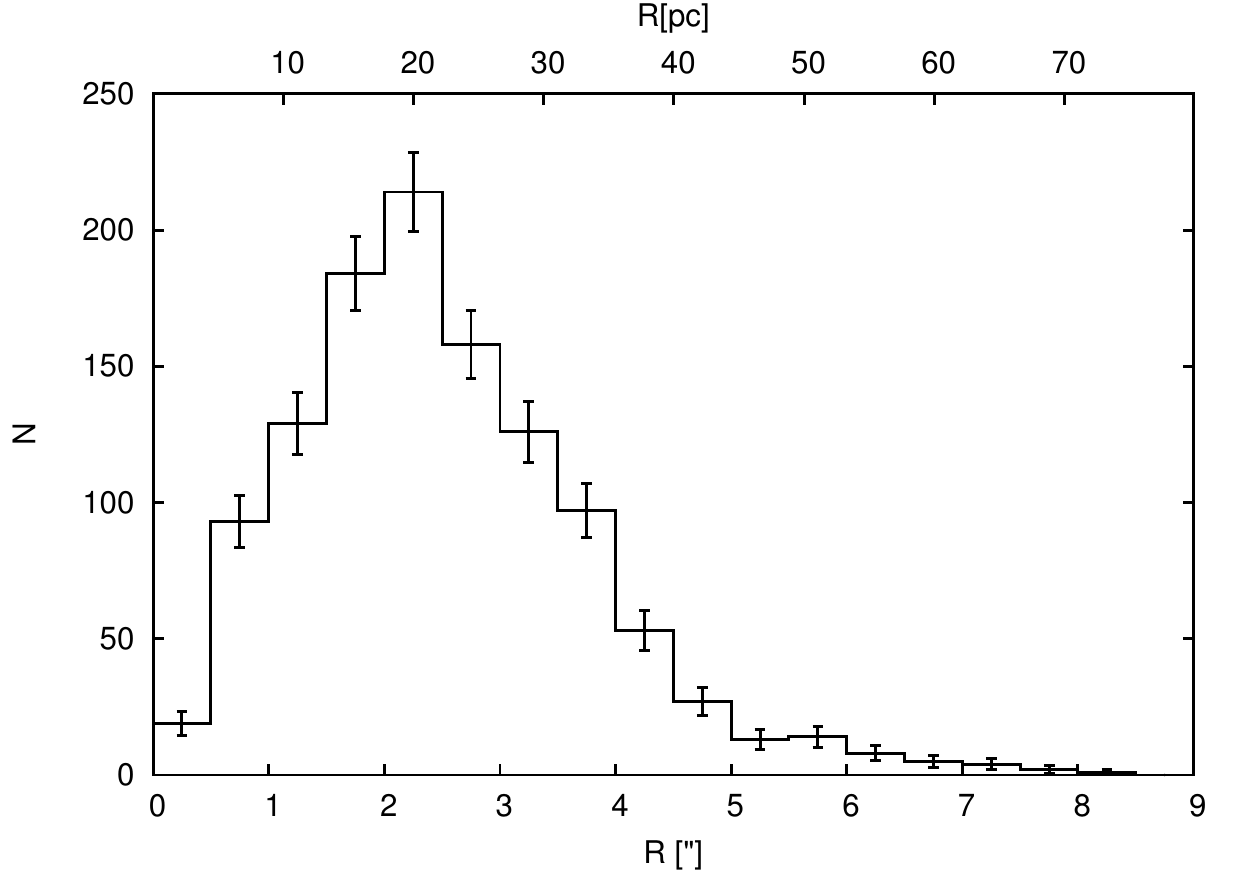}} 
\caption{Size distribution of young star groups in NGC 300.}
\label{H_radio}
\end{figure}

From the size distribution of the 1147 young star group detected using the PLC method shown in Fig. 
\ref{H_radio}, we derived that the average radius and the highest value in the distribution
are both $\sim$ 2.5 arcsec.
At the considered distance this is approximately equivalent to 25 pc.
This value is lower than the one found by \citet{2001A&A...371..497P} of $\sim$ 57 pc.
This difference is expected since the HST/ACS images used in this work have
higher resolution than the ground-based images. 
For this reason, they applied the PLC method with a parameter of distance $d_{s}$
much greater than ours (7, 7.5 and 8 arcsec). 
\citet{1998AJ....116..119B} performed a study of OB association for
seven spiral galaxies observed with the HST (NGC 925, NGC 2090, NGC 2541, NGC 3351, NGC 3621, NGC 4548,
and M101), finding that the size distribution peaked between radius of 20-40 pc, 
which is consistent with our value. \citet{2005A&A...440..783P} compared these data with several catalogs for OB associations
in nearby galaxies in the Local Group (IC 1613, LMC, M 33, NGC 6822, LMC, SMC, and M31) and some more distant galaxies 
(UGC 12732, NGC 1058, NGC 7217, NGC 4394, NGC 3507, and NGC 337A). They found a peak in the size distribution
between 25-55 pc and an average radius between 20-90 pc for these
galaxies. These values agree well with our results.

\subsection{Hierarchical structure in NGC~300}
\label{hierarchical}

\begin{figure*}
\centering
\resizebox{\hsize}{!}
{\includegraphics[scale=0.2]{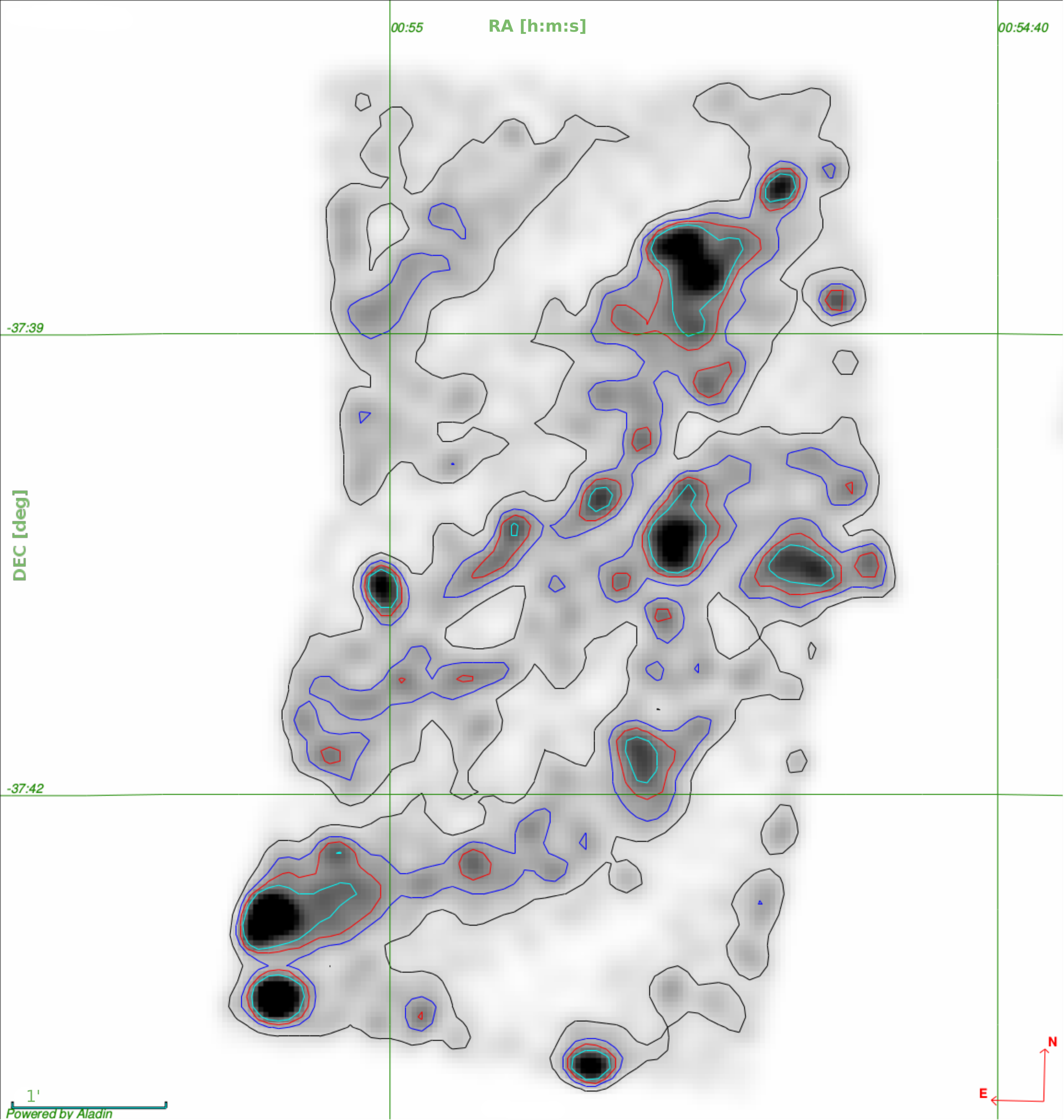}
\includegraphics[scale=0.2]{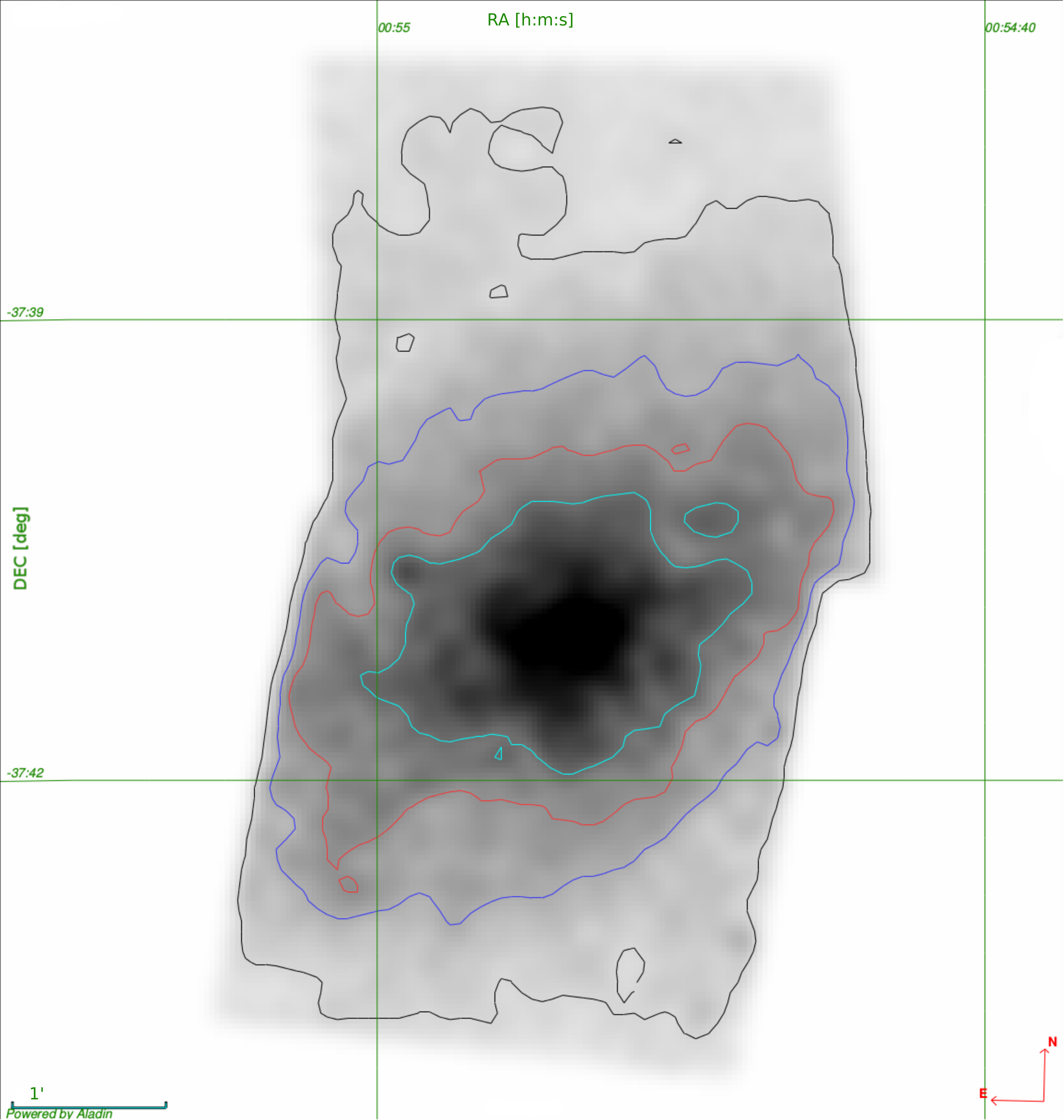}}
\caption{Density maps of central fields for the blue bright group (left panel) and the red bright group (right panel). 
The overlapping contours correspond to different pixel values, indicating different
density levels:
black 40,  blue 80, red 110,  and turquoise 145 stars per bin of 8 arcsec$^2$. }
\label{contour}
\end{figure*}

\begin{figure*}
\centering
\includegraphics[width=18cm]{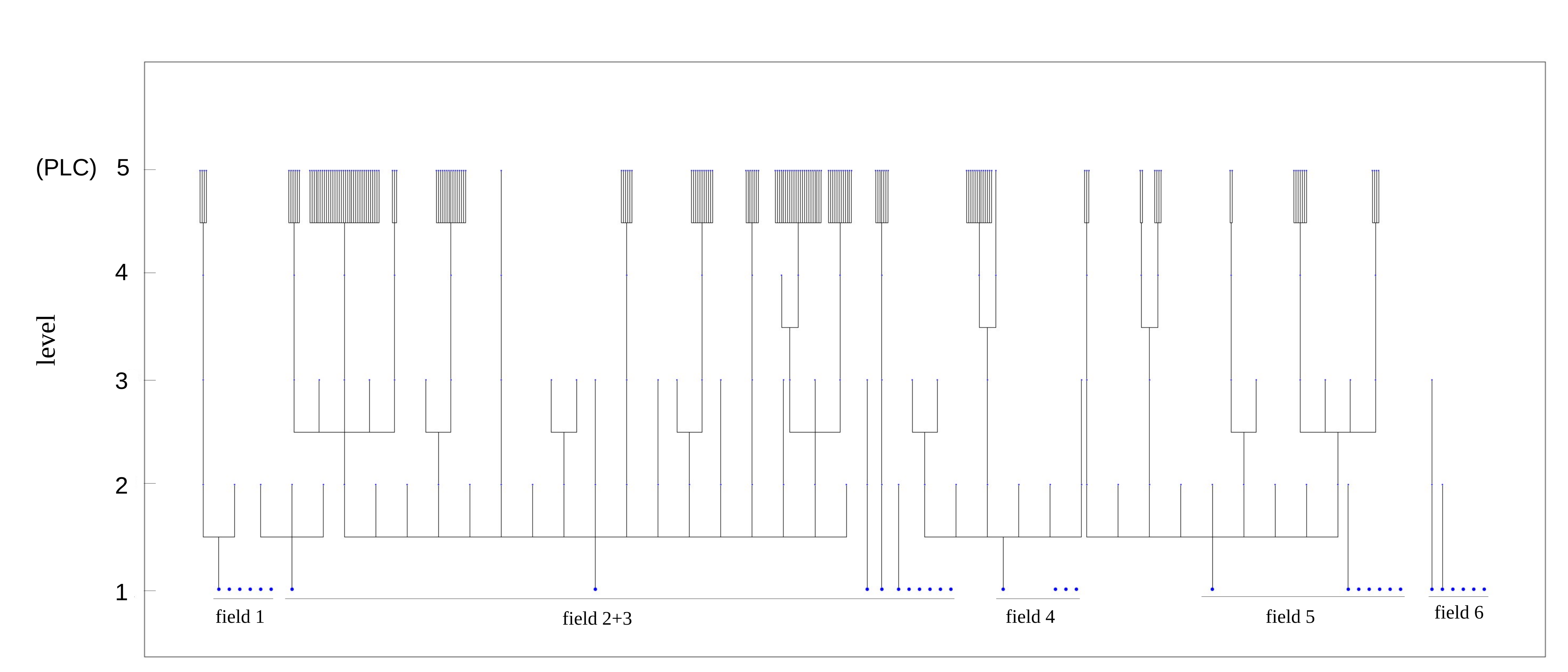}
\caption{Dendrogram of the young stellar structures detected at different density level.
Levels 1-4 correspond to a pixel value in the density maps of 40, 80, 110, and 145, respectively. Level 5
corresponds to the most compact cluster detected using PLC. The blue circles at the bottom indicate the 
structures detected at the lower density level. Some of them are no longer detected at higher densities.}
\label{dendrogram}
\end{figure*}

It is known that young stars cluster tend to be part of larger groups, such as
stellar complexes, which are grouped themselves in larger structures. 
While we searched for the best-fit distance parameter for the PLC, it became evident
that the blue stellar population in NGC~300 presents a hierarchical structure,
which starts with the spiral arms and concluds with individual compact clusters,
or even with multiple stars. 
To show this, we traced contours at different density levels on our
density stellar maps, created as described in Sect \ref{density}. In Fig. \ref{contour} 
we show the different density contours applied to the density maps of the central fields (fields 2 and 3)
for the blue bright (left panel) and the red bright group
(right panel). 
The isocontours were created using the  tool cont  of ALADIN with four different density values 
(40, 80, 110, and 145 stars per bin of 8 arcsec$^2$); this means different density levels in the density stellar maps. 
It is clear from Fig. \ref{contour} that the blue population
follows the spiral arm structure of the galaxy, while the red population is concentrated in the galaxy bulge.
The left panel shows several young structures at different density levels.
The denser smaller concentrations tend to belong to larger and looser ones, 
pointing a hierarchical structure in the distribution of blue stars in this galaxy.
To easily visualize the ties among structures of different density levels, we built a tree diagram;
these types of diagrams are called dendrograms \citep{2010ApJ...725.1717G} and indicate the connection
between parent and child structures corresponding to lower and higher density levels, respectively.
In Fig. \ref{dendrogram} we show the dendrograms of the blue stellar structures identified in the four density 
levels mentioned above. We added a last level for the most compact structures detected using PLC. 
The pixels size of the density maps is 2 arcsec, therefore it is not possible to detect 
the smallest groups that are effectively detected with PLC.

\subsection{PDMF behavior}

\begin{figure}
\resizebox{\hsize}{!}{\includegraphics{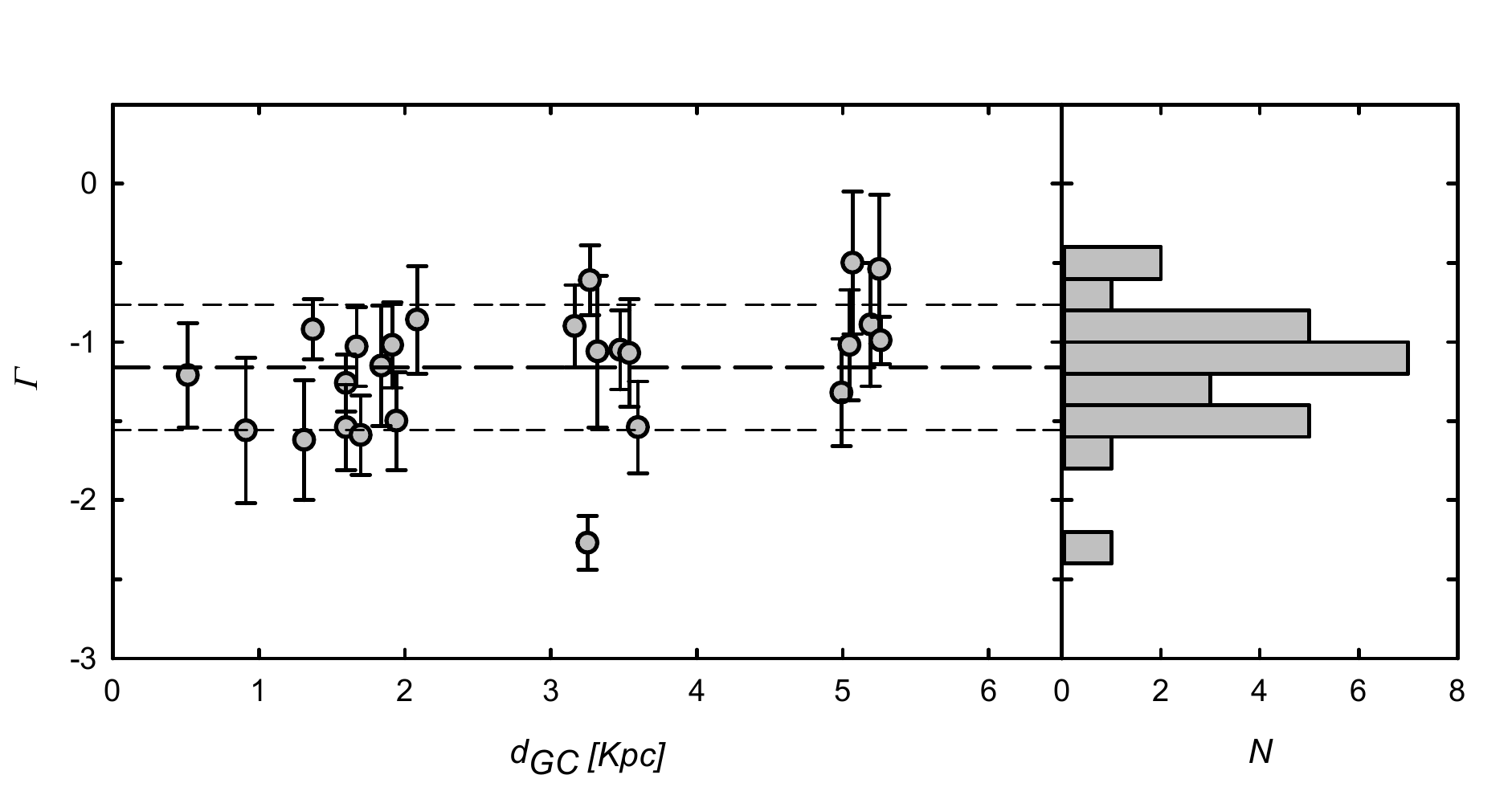}} 
\caption{PDMF behavior for the most important groups found in NGC 300 (see text for
details). The long dashed line indicates the mean slope value, and the short dashed lines
show the envelopes at one standard deviation.}
\label{dist_pend}
\end{figure}

Several works have claimed a possible relation between the 
PDMF shape and the galactocentric distance in galaxies \citep[e.g.,][]
{2015MNRAS.447.1033M}, since its trend would depend on the local 
properties within a galaxy. Searching for a possible relation of this kind 
in NGC~300, we derived the galactocentric distance of each stellar 
group considering the geometrical parameters of the galaxy given by 
\cite{1990AJ....100.1468P}. 
These are a position angle $\gamma = 114.3^{\circ}$ and an 
inclination $i = 50^{\circ}$. To obtain a more reliable result, we only 
considered the most relevant stellar groups found in NGC~300, that is, 
those with the lowest PDMF slope errors ($e_{\Gamma}< 0.5$), with 
massive stars ($M>25 M_{\odot}$), and a large number of blue stars 
($N>100$). Our sample of stellar groups was then 25, and they are plotted 
in Fig \ref{dist_pend}. It seems there is a slight trend for flatter slopes 
in the outside limit of the galaxy than in the central region (left panel).
We found a mean slope of $-1.3 \pm 0.1$ for $d_{GC}$ < 2 kpc and 
$-0.9 \pm 0.2$ for $d_{GC}$ > 4 kpc.
However, within the errors, we are unable to confirm a clear dependence of the 
$\Gamma$ slope with the galactocentric distance for NGC~300. 
We can only indicate their distribution (right panel) and their average 
and standard deviation (s.d.) values as -1.16 and 0.39, respectively.

\section{Conclusions}
\label{conclusions}

Using complementary methods for the detection of young stellar groups in NGC 300, we found that 
the blue population is located in the spiral arms of the galaxy, presenting a hierarchical behavior
where denser and smaller structures are enclosed in larger and less dense ones. We were able to delineate the hierarchical
structure on several density levels.

Through the PLC technique, 1147 young stellar groups have been detected in six fields of the galaxy NGC 300.
The resulting catalog contains the coordinates of the groups, sizes, number of stars, 
PDMF slope with its error, and galactocentric distance. In Table \ref{tabla2} we present the first ten rows. The completed
catalog is available online. The identification name of each grouping consists of four numbers: 
the first is the field number, 1, 2, 4, 5, or 6, because fields 2 and 3 were joined into a single field 
denoted by number 2 (see Sect \ref{catalog}), and the remaining three numbers indicate the number of group in the field.

\begin{table*}
\tiny
\caption{Young star groups in NGC 300}
\label{tabla2}
\centering 
\begin{tabular}{c c c c c c c c c c c c c c c c c} 
\hline\hline 
name   & field  &   raj2000    &   dej2000  &    r[$``$]&    N & N$_{bri}$ &  N$_{dct}$ &  N$_{dct-bri}$ &  N$_{blue}$  &   N$_{blue-dct}$  &   N$_{red}$ &  N$_{red-dct}$ & Mag$_{min}$ & $\Gamma$  &  err$_{\Gamma}$  &  d$_{GC}$\\
       &        &   (deg)      &    (deg)   &           &      &              &         &                   &            &                       &              &                 &     &               &      &    [Kpc] \\   
\hline
 
1001  &   1  &  13.890841 &  -37.688836   &  0.57   &    12       &       7     &        11     &         7     &        10     &        10     &         1     &         0       &       21.52      &     -0.20     &       0.31     &   4.94\\
1002  &   1  &  13.933404 & -37.685159    & 0.38    &   10        &     10      &       10      &       10      &        7      &        7      &        0      &        0        &      23.46       &      ----      &      ----      &  6.27\\
1003  &   1  &  13.890247 &  -37.688886   &  0.78   &    23       &       6     &        20     &         6     &        12     &        10     &         5     &         4       &       22.60      &     -0.00     &       0.80     &   4.93\\
1004  &   1  &  13.904136 &  -37.693602   &  1.46   &    53       &       9     &        40     &         8     &        32     &        25     &        13     &         9       &       21.09      &     -0.87     &       0.29     &   5.28\\
1005  &   1  &  13.894911 &  -37.690068   &  1.50   &    53       &       9     &        36     &         6     &        30     &        21     &        16     &        11       &       23.30      &      ----   &         ----   &   5.05\\
1006  &   1  &  13.892505 &  -37.685818   &  1.22   &    34       &       8     &        26     &         6     &        21     &        16     &         3     &         3       &       21.53      &     -0.50     &       0.42     &   5.04\\
1007  &   1  &  13.892056 &  -37.689406   &  1.00   &    33       &       6     &        26     &         5     &        18     &        14     &         8     &         6       &       22.93      &      ----     &      ----     &   4.97\\
1008  &   1  &  13.886917 &  -37.690713   &  2.44   &   153       &      14     &        89     &        11     &        77     &        46     &        46     &        25       &       21.55      &     -1.19     &       0.33     &   4.80\\
1009  &   1  &  13.889945 &  -37.697110   &  1.71   &    69       &       8     &        48     &         8     &        37     &        26     &        10     &         6       &       21.25      &     -0.29     &       0.23     &   4.82\\
1010  &   1  &  13.884190 &  -37.692673   &  1.50   &    62       &       8     &        41     &         6     &        30     &        20     &        20     &        12       &       23.80      &      ----     &      ----     &   4.70\\

\hline
\end{tabular}
\tablefoot{The suffix $bri$ indicates bright stars with $F555W < $25.
The suffix $dct$ indicate stars belonging to the decontaminated region.
The suffixes $blue$ and $red$ refer to blue and red stars respectively.
Mag$_{min}$ is the magnitude of the brightest star in the group.
$\Gamma$ is the PDMF slope.
$d_{GC}$ is the galactocentric distance.
Here we present the first ten rows, the complete table is only available in electronic form
at the CDS via anonymous ftp to cdsarc.u-strasbg.fr (130.79.128.5)
or via http://cdsweb.u-strasbg.fr/cgi-bin/qcat?J/A+A/.}
\end{table*}

The mode and the mean of the groups size distribution are both 25 pc. 
These values agree well with those found in the literature for
the Local Group and some more distant galaxies. We did not find a clear dependence of the $\Gamma$ slope 
on galactocentric distance for the PDMF of the most
important groups, but a slight trend seems to exist that deserves 
a more detailed analysis. We can only indicate that the mean value is $-1.16 
\pm 0.39$ (s.d.). 
This slope is slightly flatter than the -1.35 value given by \cite{1955ApJ...121..161S}
but is compatible within the errors.

\begin{acknowledgements} 
We thank the referee for helpful comments
and constructive suggestions that helped to improve
this paper. 
MJR and GB acknowledge support from CONICET (PIP 112-201101-00301). MJR is a fellow of
CONICET. 
Based on observations made with the NASA/ESA Hubble Space Telescope, and obtained from the 
Hubble Legacy Archive, which is a collaboration between the Space Telescope Science Institute (STScI/NASA),
the Space Telescope European Coordinating Facility (ST-ECF/ESA) and the Canadian Astronomy Data Centre (CADC/NRC/CSA).
Some of the data presented in this paper were obtained from the Mikulski Archive for Space Telescopes (MAST). 
STScI is operated by the Association of Universities for Research in Astronomy, Inc., under NASA 
contract NAS5-26555. Support for MAST for non-HST data is provided by the NASA Office of Space Science 
via grant NNX09AF08G and by other grants and contracts.
This research has made use of "Aladin sky atlas" developed at CDS, Strasbourg Observatory, France.
Finally we would like to thank to Dr. Guido Moyano Layola, the English cabinet of FCAG-UNLP, and
the language editor Astrid Peter for English corrections. 
\end{acknowledgements}

\bibliographystyle{aa} % style aa.bst
\bibliography{reference} % your references Yourfile.bib

\end{document}